\documentclass[a4paper, nofootinbib, prd, onecolumn, superscriptaddress]{revtex4-2}
\usepackage[utf8]{inputenc}
\usepackage{lipsum}
\usepackage{xcolor}
\usepackage{bm, amssymb,amsmath,amsthm,amsfonts,mathrsfs}
\usepackage{graphicx}
\usepackage{tabularx}
\usepackage{booktabs}
\usepackage[hidelinks]{hyperref}
\usepackage[capitalise]{cleveref} 
\hypersetup{
	colorlinks=true,
	linkcolor=blue,
	filecolor=blue,      
	urlcolor=blue,
	citecolor=blue
}
\usepackage{subfigure}
\usepackage[normalem]{ulem}
\newcommand{\dd}{\text{d}}

\newcommand{\be}{\begin{eqnarray}}
\newcommand{\ee}{\end{eqnarray}}

\begin{document}

\title{A Dark Matter Probe in Accreting Pulsar-Black Hole Binaries}

\author{Ali Akil}
\email{aakil@connect.ust.hk}

\affiliation{Department of Physics, Southern University of Science and Technology, Shenzhen 518055, China}
\affiliation{Department of Physics, The Hong Kong University of Science and Technology,\\
Clear Water Bay, Kowloon, Hong Kong, China}
\affiliation{Jockey Club Institute for Advanced Study, The Hong Kong University of Science and Technology,\\
Clear Water Bay, Kowloon, Hong Kong, China}

\author{Qianhang Ding}
\email{qdingab@connect.ust.hk}

\affiliation{Department of Physics, The Hong Kong University of Science and Technology,\\
Clear Water Bay, Kowloon, Hong Kong, China}
\affiliation{Jockey Club Institute for Advanced Study, The Hong Kong University of Science and Technology,\\
Clear Water Bay, Kowloon, Hong Kong, China}

\begin{abstract}

The accretion of dark matter (DM) into astrophysical black holes slowly increases their mass. The rate of this mass accretion depends on the DM model and the model parameters. If this mass accretion effect can be measured accurately enough, it is possible to rule out some DM models, and, with the sufficient technology and the help of other DM constraints, possibly confirm one model. We propose a DM probe based on accreting pulsar-black hole binaries, which provide a high-precision measurement on binary orbital phase shifts induced by DM accretion into black holes, and can help rule out DM models and study the nature of DM.

\end{abstract}

\maketitle

\section{Introduction}

Dark matter (DM) consists around $26\%$ of the energy density in the Universe \cite{Planck:2018vyg}, which deeply influences the cosmic evolution and shapes the large scale structure. With decades of observations and studies, the Lambda-cold dark matter ($\Lambda$CDM) model has become the standard model of Big Bang cosmology \cite{Rubin:1970zza, Faber:1979pp, Turner:1984nf}. However, there still exist unsolved DM problems between theoretical predications and observations, such as the core-cusp problem \cite{Flores:1994gz, Moore:1994yx, vandenBosch:1999ka}, missing satellites problem \cite{Klypin:1999uc, Moore:1999nt}, etc., which push us to study the nature of DM. 

The difficulty in studying the nature of DM is its weak or absent interaction with baryonic matter, we are not able to observe it directly. This left space for a large number of theoretical models accounting for the gravitational effects of DM, like weakly interacting massive particles (WIMPs) \cite{Chiu:1966kg, Lee:1977ua}, ultralight DM \cite{Ferreira:2020fam}, primordial black holes (PBHs) \cite{Hawking:1971ei}, modified gravity \cite{Clifton:2011jh}, etc. Some of these models have restricted parameter ranges due to observational constraints \cite{Boehm:2004th, DES:2020fxi, Carr:2020gox}. However, most of the models are not completely ruled out and are waiting to be examined by future DM probes.

To develop a DM probe to study the nature of DM, the accretion of DM into a black hole (BH) is a potential channel, where various DM models contribute different accretion rates and result in distinct BH mass increment after long enough timescale. Such an accretion effect of DM has been well studied in a number of scenarios for understanding DM properties \cite{Peirani:2008bu, Brito:2015yga, Outmezguine:2018nce, malhan2021probing, DAgostino:2022ckg}, however, due to the slow accretion rate of DM, a measurable DM accretion effect needs a supermassive host astrophysical object, which owns a complex matter surrounding environment and causes difficulty in extracting the DM information. The introduction of pulsar in BH accreting DM changes the story. The pulsar can emit stable pulse signals, which provide a high time-resolution measurement on its surrounding environment \cite{backer1982millisecond}. This high precision measurement can even measure the mass loss of a pulsar due to electromagnetic radiation \cite{Kramer:2021jcw}. When a pulsar rotates around an accreting BH, the cumulative DM accretion slowly increases the BH mass and influences orbital evolution of pulsar-black hole (PSR-BH) binaries. This deviation from a standard general-relativistic orbital evolution would produce an orbital phase shift, which can be detected in pulse timing after a long-time accretion cumulation. Since accretion effects in various DM models are different, the detected orbital phase shift induced by DM accretion is corresponding with DM models and model parameters, which can work as a DM probe in ruling out DM models and studying the nature of DM.

Although, we do not observe any PSR-BH binaries so far, numerous studies have estimated the number of PSR-BH binaries inside the Milky Way is around $\mathcal{O}(10) - \mathcal{O}(1000)$ \cite{Shao:2018qpt, Chattopadhyay:2020lff}. With the improvement of sensitivity in radio telescopes, such as Five-Hundred Metre Aperture Spherical Radio Telescope (FAST) \cite{nan2011five}, MeerKAT \cite{booth2009meerkat} and future Square Kilometre Array (SKA) \cite{dewdney2009square}, the PSR-BH binaries are expected to be observed in the near future. Also, gravitational waves (GWs) provide another window for detection. The low frequency GW detectors, like Laser Interferometer Space Antenna (LISA) \cite{amaro2017laser}, have high sensitivity in detecting GWs from small mass ratio binaries, which could be PSR-BH binaries. A joint observation by radio telescopes and GW detectors can enhance the possibility of the detection of PSR-BH binaries.

This paper is organized as follows. In Sec.~\ref{Accretion}, we give a brief introduction on the DM accretion rate into an astrophysical BH, including the accretion rate of WIMPs in Sec.~\ref{CDM}, the accretion rate of a hot particle DM model in Sec.~\ref{Relativistic}, the accretion rate of ultralight DM in Sec.~\ref{Axion}, a brief estimation on the accretion rate of PBHs in Sec.~\ref{PBH}, and a comment on baryonic matter accretion about how and in what cases it affects our proposal in Sec.~\ref{Baryonic}. Then in Sec.~\ref{Binaries}, we discuss about the pulse timing in PSR-BH binaries, including using PSR-BH binaries to study the new phenomena in Sec.~\ref{MassFixed} and studying BH mass changing effect in PSR-BH binaries in Sec.~\ref{MassChange}. Finally, in Sec.~\ref{acc_in_bin}, we numerically calculate the mass accretion effect in PSR-BH binaries and use it to constrain the DM models, including constraining WIMPs in Sec.~\ref{wimp_acc_in_bin}, ultralight DM in Sec.~\ref{uldm_acc_in_bin}, PBHs in Sec.~\ref{PBH_acc_in_bin}.

\section{Dark Matter Accretion} \label{Accretion}

Accretion into stellar objects, particularly BHs, has been studied for a long time. In 1952, the Bondi accretion formula was derived for a non-relativistic gas cloud \cite{Bondi:1952ni}. Then in the framework of General Relativity, Michel derived a formula for hot gas accretion \cite{Michel:1972}. Unruh, on the other hand, worked out the case for scalar fields by solving the Klein-Gordon equation in a Schwarzschild BH background geometry \cite{Unruh:1976fm}. The reader will notice that different DM models obey different accretion formulas, which will lead to different accretion rates. For example, WIMPs obey Bondi's formula, whereas hot particle DM, being relativistic, will obey Michel's formula, etc. In each of the coming subsections we will briefly introduce one accretion formula. All the accretion rates basically scale like the square of the BH mass $M_{B}^2$, but the constant which multiplies it varies heavily from one DM model to another. And the accreting BH that we are considering is a Schwarzschild BH.

\subsection{Weakly Interacting Massive Particles} \label{CDM} 

WIMPs were until recently a long time most popular DM candidate. WIMPs are beyond standard model particles that interact only very weakly except for their gravitational field. WIMPs mass can be anywhere between $2\,\mathrm{GeV}$ and $100\,\mathrm{TeV}$ \cite{Roszkowski:2017nbc}. The accretion of WIMPs by BHs is captured by the well known Bondi accretion formula \cite{Bondi:1952ni}. The Bondi accretion assumes a spherically symmetric stellar object stationarily accreting a cloud of non-relativistic matter particles. For practical reasons, we will refer to the form in which it is presented in \cite{Aguayo-Ortiz:2021jzv} (for a detailed derivation, see Appendix.~\ref{WIMPAppendix}), where the equation reads,
\be \label{eq:bondi_formula}
\frac{\dd M_{\rm B}}{\dd t} =4 \pi \lambda_B (G M_{\rm B} )^2 \frac{\rho_\infty  }  { \gamma^{\frac{3}{2}} \, \Theta_\infty ^{\frac{3}{2} } c^3 }~,
\ee
with
\be  
\Theta = \frac{k_B T}{m c^2} = \frac{c_s^2}{\gamma c^2}~,
\ee
being the dimensionless temperature. Here, the physical constants $G$, $c$, $k_B$ are Newton's constant, the speed of light, and the Boltzmann constant, respectively. $\rho_\infty$ is the density at infinity. Infinity in this context being what is relevant for such an astrophysical scale rather than for example cosmological scale. $M_{\rm B}$ is the mass of the accreting BH, $m$ is the WIMPs mass and $\gamma$ is the so called polytropic constant, characterizing a polytropic fluid of pressure $P$ and density $\rho$ for which $P \sim \rho^\gamma$. Moreover, $\lambda_B =\frac{1}{4} \left(\frac{2}{5 - 3 \gamma} \right)^\frac{5 - 3 \gamma}{2 (\gamma - 1)}$. $c_s$ is the sound speed of DM \footnote{An ideal CDM model is collisionless with a zero sound speed, however, CDM particles still have a non-zero velocity dispersion (see \cite{Armendariz-Picon:2013jej} for details), which causes a free streaming away from gravitational collapse and produces an effective sound speed.}, and it is constrained by the rotation curve of the Milky Way as $c_s < 10^{-4} c$ \cite{Avelino:2015dwa}, which puts an upper bound on dimensionless temperature $\Theta < \mathcal{O}(10^{-8})$. As one would expect, the colder (smaller dispersion) the DM is, the higher the accretion rate. Moreover, the heavier it is (within the range of validity of the Bondi accretion), the higher its accretion rate. 

\subsection{Hot Particle Dark Matter} \label{Relativistic}
For the hot DM case, general relativistic effects are quite important, therefore we deal with Lorentz covariant quantities and have a full general relativistic treatment. This is captured by the Michel accretion formula \cite{Michel:1972}, which has the same form as Bondi's, with a different $\lambda_M \neq \lambda_B$, a smaller mass and higher temperature, causing a much smaller accretion rate (for detailed derivation, see Appendix.~\ref{HotAppendix}).
\be \label{Michel}
\frac{\dd M_{\rm B}}{\dd t} = 4 \pi  \lambda_M (G M_{\rm B} )^2 \frac{\rho_\infty  }  { \gamma^{\frac{3}{2}} \, \Theta_\infty ^{\frac{3}{2} } c^3 }~.
\ee
Where $\lambda_M$, unlike in the Bondi accretion, depends also on the sound speed of the medium but in general varies between $1$ and $2$ \cite{Aguayo-Ortiz:2021jzv}. Since this accretion rate is extremely small and cannot be observed in the method we propose, we will not elaborate on it. But it is good to note that this model is one of those that will be favoured by the experiment that we propose in the case of the measurement showing there is no growth in the  astrophysical BH mass.

\subsection{Ultralight Dark Matter} \label{Axion}

Recently, the scalar field models of DM have arguably become the most popular. The ultralight DM model, on the small mass between $10^{-24}\,{\rm eV}$ and $1 \, {\rm eV}$, is believed to solve the small scale problems of CDM \cite{Ferreira:2020fam}. 

The accretion rate of the ultralight DM is derived from the Klein-Gordon equation on a non-rotating BH background \cite{Unruh:1976fm}. The mass accretion rate by a non-rotating BH of mass $M_\mathrm{B}$ traveling with velocity $v$ through uniformly distributed ultralight DM of mass $m_\mathrm{ul}$ and density $\rho_\mathrm{DM}$ can be expressed as follows,
\begin{align}\label{eq:ultralightDM}
	\frac{\dd M_\mathrm{B}}{\dd t} = \frac{32 \pi^2 (GM_\mathrm{B})^3 m_\mathrm{ul} \rho_\mathrm{DM}}{\hbar c^3 v [1-\exp(-\xi)]}~,
\end{align}
where $\xi$ is defined as $\xi \equiv 2 \pi G M_\mathrm{B} m_\mathrm{ul}/\hbar v$ and $\hbar$ is the reduced Planck's constant.

\subsection{Primordial Black Holes} \label{PBH}

Apart from ultralight DM with mass smaller than $1\,\mathrm{eV}$ and WIMPs with mass between $2\,\mathrm{GeV}$ and $100\,\mathrm{TeV}$, some ultraheavy objects can also be the DM candidate, in particular, primordial black holes (PBHs). PBHs were first introduced in \cite{Hawking:1971ei}, which proposed that primordial perturbations could collapse to a BH, and PBHs was later used in accounting for massive astrophysical compact halo objects (MACHOs) \cite{Kawasaki:1997ju, Jedamzik:1998hc}.

As PBHs are much heavier than the other DM models that we have considered above, the way they can be accreted by astrophysical BHs differs radically. Consider the mass of PBH is above $1\,M_\odot$, the accretion of a PBH in an astrophysical BH behaves like a binary evolution, and its accretion rate is related with the merger rate of this binary system, which can be estimated with the mean free path of the astrophysical BH and its moving velocity. 

Given the amount of DM in our galaxy, if we distribute it into PBHs rather than the other types of DM, those would be highly separated from each other, due to their large mass. Here, we assume the PBH mass function is monochromatic. Then, the mean free path $l$ in this system depends on the cross section of BHs $\sigma \simeq 27 \pi (GM_\mathrm{B}/c^2)^2$ \cite{zakharov1988effective} and number density of PBHs $n = \rho_\mathrm{DM}/M_\mathrm{PBH}$, which can be expressed as follows,
\begin{align}\label{eq:mean_free_path}
	l_f = \frac{1}{\sigma n} = \frac{1}{27 \pi} \left( \frac{c^2}{G M_\mathrm{B}} \right)^2 \frac{M_\mathrm{PBH}}{\rho_\mathrm{DM}}~.
\end{align}
Then, the mean free time for this astrophysical BH can be approximated as $t_f \sim l_f/v$, where $v$ is the relative velocity between the astrophysical BH and PBHs. The PBHs accretion rate can be evaluated as follows,
\begin{align}\label{eq:PBH_acc}
	\frac{\dd M_\mathrm{B}}{\dd t} \simeq \frac{M_\mathrm{PBH}}{t_f} \simeq 27 \pi  (G M_\mathrm{B})^2 \frac{\rho_\mathrm{DM} v}{c^4}~.
\end{align}

\subsection{Comment on Baryonic Matter Accretion} \label{Baryonic}

One important thing to take into account is the accretion of the baryonic matter. A detailed and careful account is certainly needed. However, as we keep the precise computation for a future project, we try here discuss the magnitude of that accretion. First, one should notice that if the PSR-BH binary is found outside the galactic disk, the baryonic matter there is very scarce and baryonic accretion is to be comfortably ignored.  On the other hand, if the binary is in the galactic disk, we might expect a significant baryonic matter accretion. However, as we are comparing the behaviour of the binary for different DM models, and that baryonic matter accretion is the same in both cases, this will decrease its importance in the process. It will contribute as an added term to the BH mass, in both the cases that we will be comparing, which decreases its significance. And finally, an important point is that in our galaxy, the baryonic matter is in its great majority in stars (around $90\%$, see \cite{binney2011galactic} page 2). Therefore, we assume that if a whole star is swallowed by a Milky Way BH we will be able to see that.

\section{Pulsar-Black Hole Binaries} \label{Binaries}

\subsection{Background} \label{MassFixed}

The PSR-BH binary system is the holy grail in radio astronomy, since the stable pulse signals emitted from pulsar can provide high precision measurements on the strong gravity field around the BH \cite{Liu:2014uka}. Due to the motion of the pulsar around the BH, the pulse signals could be affected by the periodic motion of the pulsar and the gravitational field in the binary, and such effects contribute three different kinds of time delay in measuring the orbital Time-of-Arrival (TOA) of receiving pulse signals \cite{Taylor:1994zz} as follows,
\begin{align}
	\Delta_\mathrm{orb}\mathrm{TOA} = \Delta_R + \Delta_E + \Delta_S~.
\end{align}
Here, $\Delta_R$ is the R{\o}mer delay, which describes the elapsed time of light in passing the binary orbit. $\Delta_E$ is the Einstein delay that is the general-relativistic time dilation in the PSR-BH gravitational field. $\Delta_S$ is the Shapiro delay, which is the delayed time of light in curved spacetime. 

After measuring TOA of receiving pulses, we can use them to fit a given model by minimizing the timing residual between data and model predictions (see TEMPO2 program \cite{Hobbs:2006cd} for details). It gives five Post-Keplerian (PK) parameters, which can help determine the mass of the binary components and the orbital parameters of the binary. With these PK parameters, new phenomena can be tested in a given binary model. The new phenomena produces a different gravitational effect, which could influence the orbit evolution in a binary, and such an orbit deviation would produce a significant orbital phase shift, which could be detected after long enough observation time. In the general relativistic background, the orbital phase shift $\Delta \phi$ can be calculated as follows,

\begin{align}\label{eq:phase}
	\Delta \phi(t) = 2 \pi\int_0^t f(\tau) d \tau - 2 \pi \int_0^t f_\mathrm{GR}(\tau) d \tau ~,
\end{align}
where $f_\mathrm{GR}$  is the orbital frequency in general relativity without new phenomena and $f$ is the orbital frequency with new phenomena.

In order to ensure the detection of orbital phase shift induced by the new phenomena, the measurement uncertainty of the orbital phase shift $\sigma_{\Delta \phi}$ should be smaller than the measured $\Delta \phi$. The uncertainty of orbital phase shift is determined by the single measurement error and the number of independent measurements. Assuming the observation time per day $t_\mathrm{obs} \simeq 10 \, \mathrm{hrs}$, the orbital phase error in a single continuous measurement is decided by the orbital phase measurement error within one orbital period $\sigma_\phi$ divided by the number of orbital periods within one continuous observation $N_{P_\mathrm{b}}$,  which is $\sigma_\phi/N_{P_\mathrm{b}}$. Here $\sigma_\phi$ can be maximally estimated as a pulse period $P$ divided by orbital period $P_\mathrm{b}$ and $N_{P_\mathrm{b}}$ equals $t_\mathrm{obs}/P_\mathrm{b}$. In calculating the number of independent measurements, we assume the observation is under running once per day, the number of independent measurements within observation time $t$ is $t/1 \, \mathrm{days}$. Then the uncertainty of the orbital phase shift can be calculated as follows (also see \cite{Ding:2020bnl, Tong:2021whq}),
\begin{align}\label{eq:uncertainty}
	\sigma_{\Delta \phi} = \frac{2 \pi}{\sqrt{t/1 \, \mathrm{day}}}\frac{P}{t_\mathrm{obs}}~.
\end{align}
The detection of new phenomena in PSR-BH binaries requires two conditions, one is that the orbital phase shift within observation time should be larger than its measurement uncertainty and the other one is that the observation time for the orbital phase shift cannot be longer than the duty time of the radio telescope $T_\mathrm{duty}$ and the merger time of the binary $T_\mathrm{merger}$. The conditions can be expressed as follows,
\begin{align}\label{eq:require}
	|\Delta \phi(t)| > \sigma_{\Delta \phi}(t)~,~~~t < \min(T_\mathrm{duty}, T_\mathrm{merger})~.
\end{align}

\subsection{Mass Changing Binaries} \label{MassChange}

The physical background of the PSR-BH binary is very complex, it includes the baryonic matter and DM. Such a matter surrounding environment could cause matter accretion effect around the BH, which slowly increases the mass of the BH (the accretion effect around the pulsar is neglected, due to its relatively small mass, the ratio of accretion rate between pulsar and BH would be smaller than $\mathcal{O}(1\%)$, if the BH mass is larger than $10 \, M_\odot$). Although this matter accretion effect is extremely weak, it can still induce a detectable orbital phase shift in PSR-BH binaries after a long enough observation time. This orbital phase shift depends on the surrounding matter density and matter properties, especially, DM properties. Various DM models and model parameters would produce different orbital phase shifts in the PSR-BH binary, which can be used to distinguish the DM models and constrain the parameter regions of DM within a model.

In this scenario, studying the orbital evolution of the PSR-BH binary with a changing BH mass is essential. During the orbital evolution of the PSR-BH binary, the gravitational radiation takes the gravitational energy and angular momentum away from the binary system. Follow \cite{Peters:1964zz} which is calculated as follows,
\begin{align}\label{eq:GW_rad}\nonumber
	&P = \frac{G}{5 c^5} \left( \frac{\dd^3 Q_{ij}}{\dd t^3}\frac{\dd^3 Q_{ij}}{\dd t^3} - \frac{1}{3}\frac{\dd^3 Q_{ii}}{\dd t^3}\frac{\dd^3 Q_{jj}}{\dd t^3} \right)~,\\
	&\frac{\dd L_i}{\dd t} = -\frac{2 G}{5 c^2}\epsilon_{ijk}\frac{\dd^2 Q_{mj}}{\dd t^2}\frac{\dd^3 Q_{mk}}{\dd t^3}~.
\end{align}
Here, $G$ is the Newton's constant, $c$ is the speed of light, $\epsilon_{ijk}$ is the three-dimensional Levi-Civita symbol, $Q_{ij}$ is a tensor which is defined as $Q_{ij} = \sum_{\alpha} m_\alpha x_{\alpha i} x_{\alpha j}$, its form in a binary system can be expressed as $Q_{xx} = \mu d^2 \cos^2\phi$, $Q_{yy} = \mu d^2 \sin^2\phi$ and $Q_{xy} = Q_{yx} = \mu d^2 \sin\phi\cos\phi$, where $\mu$ is the reduced mass $m_1 m_2/(m_1 + m_2)$, $d$ is the distance between components in the binary and $\phi$ is the orbital phase of the binary.

In Eq.~\eqref{eq:GW_rad}, the magnitude of the gravitational radiation is determined by the time derivative of $Q_{ij}$, where a changing BH mass contributes. Such a changing BH mass not only contributes to the gravitational radiation, but also influences the gravitational potential energy of the binary. Due to the accretion of DM into BHs, the gravitational potential energy is transferred from the DM to the PSR-BH binary, an overall contribution of gravitational potential energy can be estimated as follows, a detailed derivation can be found in Appendix.~\ref{GPAppendix}.
\begin{align}\label{eq:GP}
	\frac{\dd E_{p}}{\dd t} = -\frac{G m_\mathrm{p}}{a}\frac{\dd M_\mathrm{B}}{\dd t}~,
\end{align}
where $a$ is the semi-major axis of the binary, $m_\mathrm{p}$ is the mass of the pulsar and $M_\mathrm{B}$ is the BH mass. The total energy $E$ and angular momentum $L$ of the PSR-BH binary are related with their orbital parameters, semi-major axis $a$ and eccentricity $e$ as follows \cite{Peters:1964zz},
\begin{align}\label{eq:ELae}
	a = -\frac{G m_\mathrm{p} M_\mathrm{B}}{2E}~,~~~L^2 = \frac{G m_\mathrm{p}^2 M_\mathrm{B}^2}{m_\mathrm{p} + M_\mathrm{B}} a (1-e^2)~.
\end{align}
Combining Eqs.~(\ref{eq:GW_rad}--\ref{eq:ELae}), the time derivative of orbital parameters $da/dt$ and $de/dt$ can be numerically solved from the energy conservation and angular momentum conservation, which can be expressed as follows,
\begin{align}
	\frac{\dd E}{\dd t} = -P_\mathrm{acc} + \frac{\dd E_p}{\dd t}~,~~~\frac{\dd L}{\dd t} = \frac{\dd L_\mathrm{acc}}{\dd t}~.
\end{align}
Here, $P_\mathrm{acc}$ is the power of the gravitational radiation emitted from the PSR-BH binary with a DM accretion into the BH. $\dd L_\mathrm{acc}/ \dd t$ is the time derivative of angular momentum with the DM accretion effect. After obtaining the time derivative of the semi-major axis $\dd a/\dd t$, the time derivative of orbital frequency $\dd f/\dd t$ can be calculated from the Kepler's third law as follows,
\begin{align}\label{eq:fre_acc}
	\frac{\dd f}{\dd t} = \frac{1}{4 \pi}\frac{ a^{-5/2} G^{1/2}}{(m_\mathrm{p}+M_\mathrm{B})^{1/2}}\left(a \frac{\dd M_\mathrm{B}}{\dd t}-3(m_\mathrm{p}+M_\mathrm{B})\frac{\dd a}{\dd t}\right)~.
\end{align}
Meanwhile, the orbital frequency without the DM accretion follows the standard general relativistic evolution, which gives its time derivative as follows,
\begin{align}\label{eq:fre_GR}
	\frac{\dd f_\mathrm{GR}}{\dd t} = -\frac{3}{4 \pi} \frac{G^{1/2}(m_\mathrm{p} + M_\mathrm{B})^{1/2}}{a_\mathrm{GR}^{5/2}}\frac{\dd a_\mathrm{GR}}{\dd t}~.
\end{align}
Here, subscript GR denotes that the evolution of physical quantity follows the standard general relativitic evolution in \cite{Peters:1964zz}. Then the corresponding frequency evolution $f(t)$ and $f_\mathrm{GR}(t)$ can be numerically solved, following Eq.~\eqref{eq:phase}, orbital phase shift $\Delta \phi$ induced by the DM accretion on the BH can be obtained. In order to make sure a detection of DM accretion into the BH, Eq.~\eqref{eq:require} needs to be satisfied.

\section{Dark Matter Accretion in PSR-BH Binaries}\label{acc_in_bin}

\subsection{Weakly Interacting Massive Particles}\label{wimp_acc_in_bin}

For WIMPs, we use Eq.~\eqref{eq:bondi_formula} to calculate the effect of its accretion on PSR-BH binaries. It shows that the DM accretion rate into BHs depends on the DM density, BH mass and dimensionless temperature. Since the measurable pulsar systems live in the Milky Way, the DM density follows a galactic DM density profile, which we use the Navarro-Frenk-White (NFW) model to describe \cite{Navarro:1995iw, Navarro:1996gj},
\begin{align}
	\rho(r) = \frac{\rho_0}{\frac{r}{r_0}(1 + \frac{r}{r_0})^2}~,
\end{align}
where $\rho_0$ is the characteristic density and $r_0$ is the scale length. By fitting the rotation curve of the Milky Way, we have the best fit parameters are $\rho_0 = 0.052 \, M_\odot/\mathrm{pc}^3$, $r_0 = 8.1 \, \mathrm{kpc}$ \cite{Lin:2019yux}.

\begin{figure}[htbp] \centering
	\includegraphics[width=8cm]{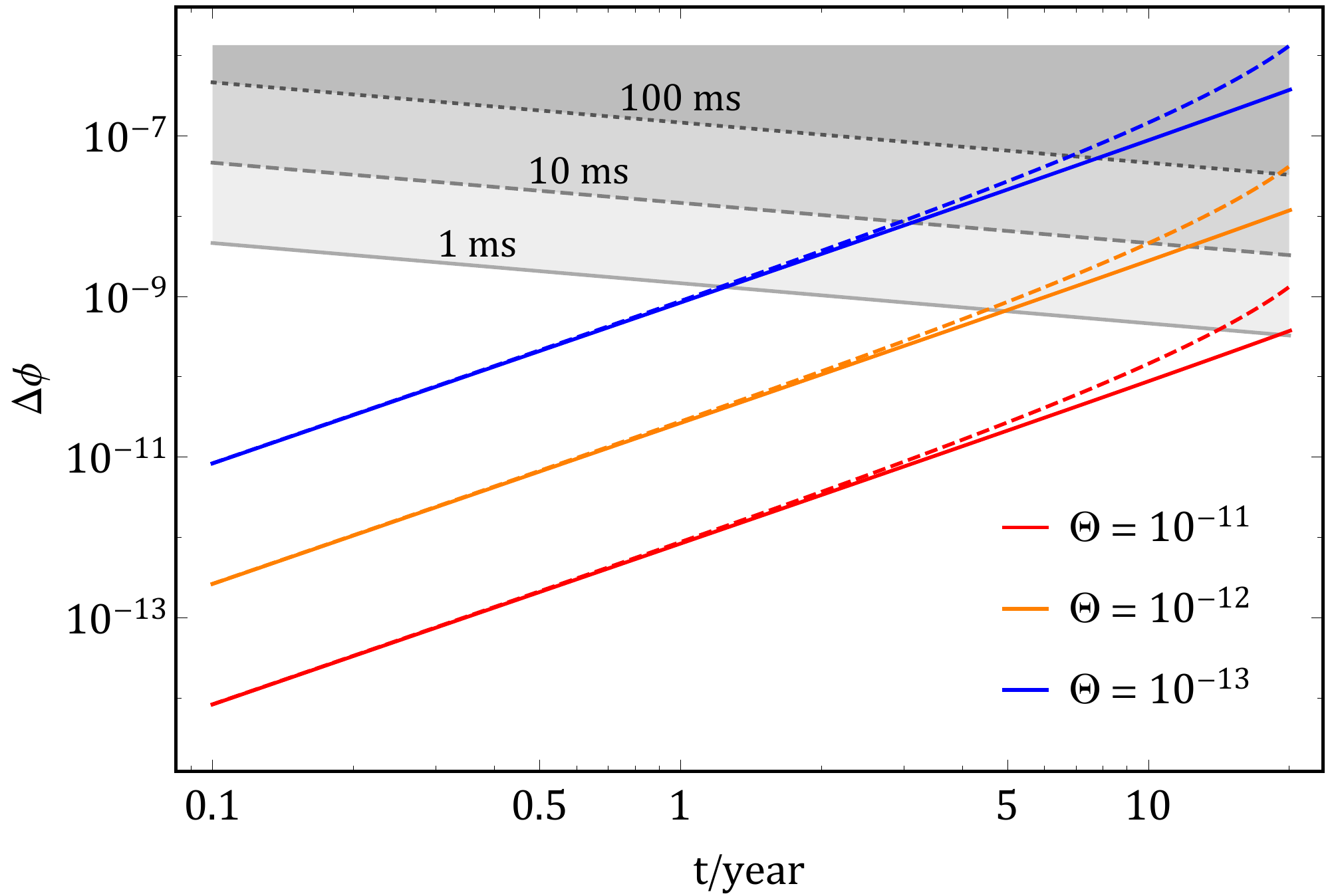}
	\caption{\label{fig:orbital_phase}
		The evolution of orbital phase shift in the PSR-BH binary. We assume that the mass of BH is $100 \, M_\odot$, the mass of pulsar is $1.6 \, M_\odot$, the initial detected GW frequency is $0.01 \, \mathrm{Hz}$ and the eccentricity is $e = 0$ ($e = 0.6$) for solid curves (dashed curves). The gray shadow regions are detectable orbital phase shift ranges by pulsar with pulse period $1 \, \mathrm{ms}$, $10 \, \mathrm{ms}$ and $100 \, \mathrm{ms}$, respectively. The different colors of curves denote dimensionless temperature of WIMPs are $10^{-11}$, $10^{-12}$ and $10^{-13}$, respectively.
	}
\end{figure}

In a practical calculation, we take the position of the PSR-BH binary to be at $10\,\mathrm{kpc}$ away from the center of the Milky Way. Following the process introduced in Sec.~\ref{Binaries}, the evolution of the orbital phase shift can be obtained, as shown in Fig.~\ref{fig:orbital_phase}.
We can find that a detectable timescale for WIMPs ($\Theta \sim 10^{-12}$) accretion on $100 \, M_\odot$ BHs in the PSR-BH binary system is $\mathcal{O}(10) \, \mathrm{years}$. A smaller $\Theta$ represent a lower temperature of WIMPs or heavier WIMPs mass, which could effectively increase the accretion rate in Eq.~\eqref{eq:bondi_formula} and enlarge the orbital phase shift in PSR-BH binaries. Moreover, a small pulse period would effectively decrease the detection timescale, due to the small measurement uncertainty in Eq.~\eqref{eq:uncertainty}, which indicates that millisecond PSR-BH binaries could be a potential candidate for studying the nature of dark matter. Also, millisecond pulsars are expected to form binaries with BHs via the ``recycle'' process \cite{Clausen:2014ksa}.

\begin{figure}[htbp] \centering
	\includegraphics[width=8cm]{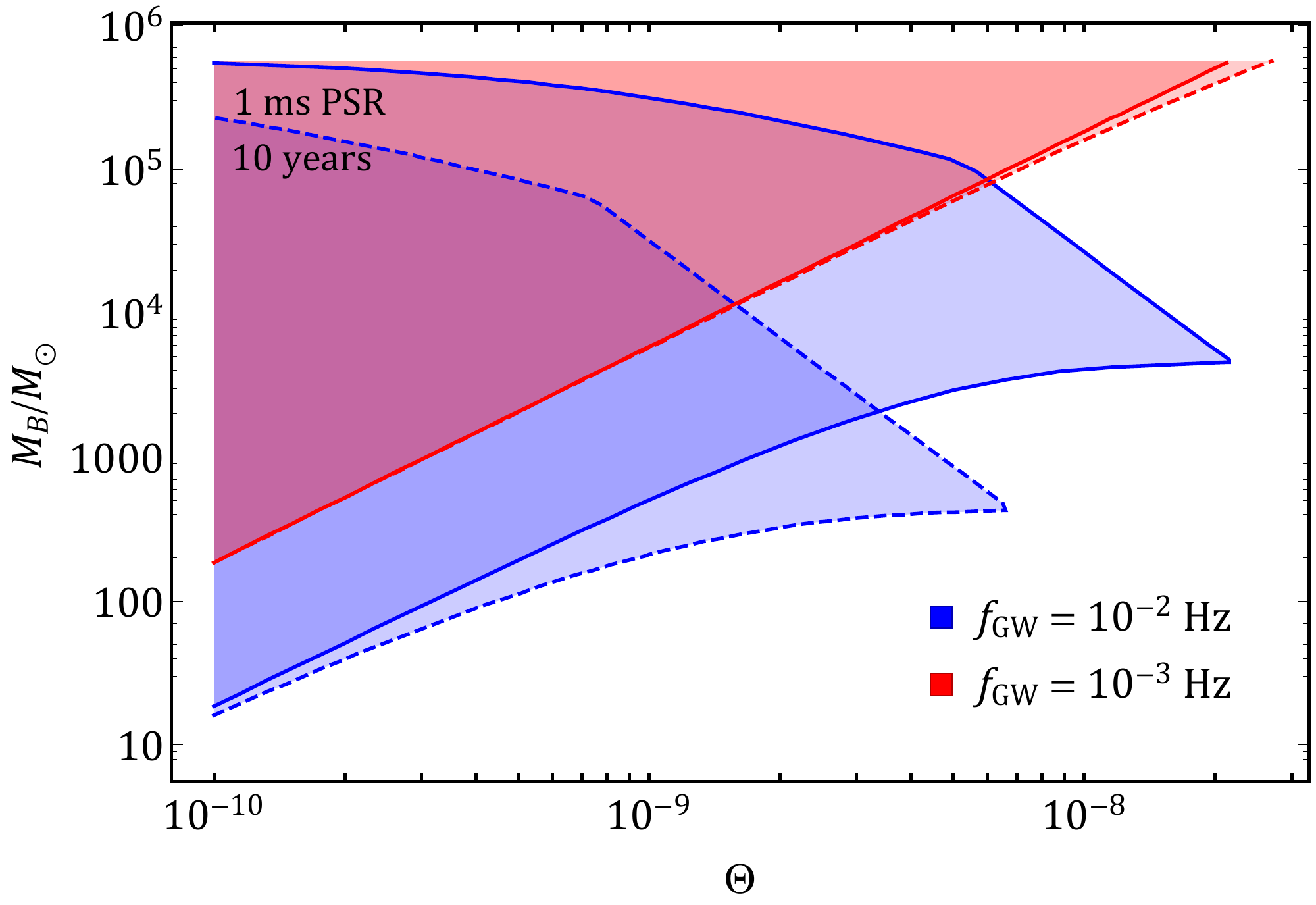}
	\caption{\label{fig:mass_theta}
		The detectable regions for dimensionless temperature $\Theta$ with different BH masses in PSR-BH binaries. We assume the pulsar mass is $1.6 \, M_\odot$, pulse period is $1 \, \mathrm{ms}$ and the duty time of radio telescope is $10 \, \mathrm{years}$. The blue (red) shadow regions represent the result from PSR-BH binaries with GW frequency $10^{-2} \, \mathrm{Hz}$ ($10^{-3} \, \mathrm{Hz}$). The solid (dashed) regions represent the result from PSR-BH binaries with eccentricity $e = 0$ ($e = 0.6$).
	}
\end{figure}

In order to obtain a detectable parameter region for the BH mass and dimensionless temperature, we follow the Eq.~\eqref{eq:require} in constraining parameters, where we assume the pulsar mass is $1.6 \, M_\odot$, pulse period is $1 \, \mathrm{ms}$ and the duty time of radio telescope is set as $10 \, \mathrm{years}$. The result is shown in Fig.~\ref{fig:mass_theta}.
We can find that a larger BH mass and higher orbital frequency could extend the range of a detectable dimensionless temperature. This is because that a larger BH mass and higher orbital frequency can increase the power of GW radiation, which speeds up the shrinkage of orbit of the PSR-BH binary. Then, orbital phase shift induced by the WIMPs accretion on BH would be enlarged. Also, for a larger eccentricity (dashed curves in Fig.~\ref{fig:mass_theta}), stronger GW is emitted which increases the orbital phase shift and extends the observable $\Theta$ range. An optimal detectable range of dark matter properties depends on various factors, namely, BH mass, orbital period, eccentricity, it means the most advantageous binary parameters for studying the nature of dark matter are different for various BH masses. The lower bound of shadow regions are given by the condition $\Delta \phi(T_\mathrm{duty}) = \sigma_{\Delta \phi}(T_\mathrm{duty})$, while the upper bound are given by the condition $\Delta \phi(T_\mathrm{merger}) = \sigma_{\Delta \phi}(T_\mathrm{merger})$. Therefore, we can find that even though a large BH mass could improve the detectable range of $\Theta$, its short merger time prohibits a large cumulative orbital phase and decreases the detectable range of $\Theta$. 

However, in usual recycle processes of millisecond pulsars, the mass of BH companion can hardly exceed $100 \, M_\odot$ \cite{Clausen:2014ksa, Chattopadhyay:2020lff} and such a stellar mass BH companion can only help constrain a small part of DM parameter range as shown in Fig.~\ref{fig:mass_theta}. In order to probe a larger DM parameter range, a massive BH companion (especially around $10^{4} \, M_\odot$ in Fig.~\ref{fig:mass_theta}) is needed in the PSR-BH binary, which could be formed via several channels. The first channel is having a millisecond PSR-BH binary in a dense stellar environment, then through sequences of stellar interactions, where the BH encounters and merges with a number of stars, and turns into an intermediate-mass black hole (IMBH). Finally this millisecond PSR-BH binary becomes a millisecond PSR-IMBH binary, \cite{Merritt:2011ve, Fragione:2017blf}. The second channel is having an IMBH formed near the center of a Milky Way-like galaxy or globular clusters, then the strong gravitational field of IMBH can help capture the single millisecond pulsar or millisecond pulsar binaries. As a large population of pulsars, including millisecond pulsar, is expected in globular clusters, such processes could finally lead to the formation of millisecond PSR-IMBH binaries \cite{Devecchi:2007tn, Zhang:2014kva}. The third possible alternative channel is through the so-called hierarchical triple systems. The hierarchical triple systems consist of a binary system (a pulsar and a low-mass companion) orbiting an IMBH. The tidal interactions between the inner IMBH and outer binary can disrupt the outer binary and kick out the low-mass companion, forming a tight millisecond PSR-IMBH binary \cite{toonen2016evolution}.

In this numerical calculation on the orbital phase shift, we use Eq.~\eqref{eq:fre_acc} and \eqref{eq:fre_GR}, which is the orbital evolution in non-relativistic limit. This limit is only valid in the inspiral phase of the PSR-BH binary, so an upper bound of the orbital frequency during inspiral phase should be put in calculating the orbital phase shift. This maximal inspiral phase frequency can be estimated as $f_\mathrm{insp} = (a \eta^2 + b \eta + c)/2 \pi G M$ \cite{Chernoff:1993th, Ajith:2009bn, Zhu:2011bd}. $\eta$ is the symmetric mass ratio, which is defined as $\eta \equiv m_\mathrm{p} M_\mathrm{B}/M^2$ in the PSR-BH binary, $M \equiv m_\mathrm{p} + M_\mathrm{B}$ and the coefficients $a = 0.29740$, $b = 0.04481$, $c = 0.09556$ (see Table 1 of \cite{Ajith:2007kx}).

\begin{figure}[htbp] \centering
	\includegraphics[width=8cm]{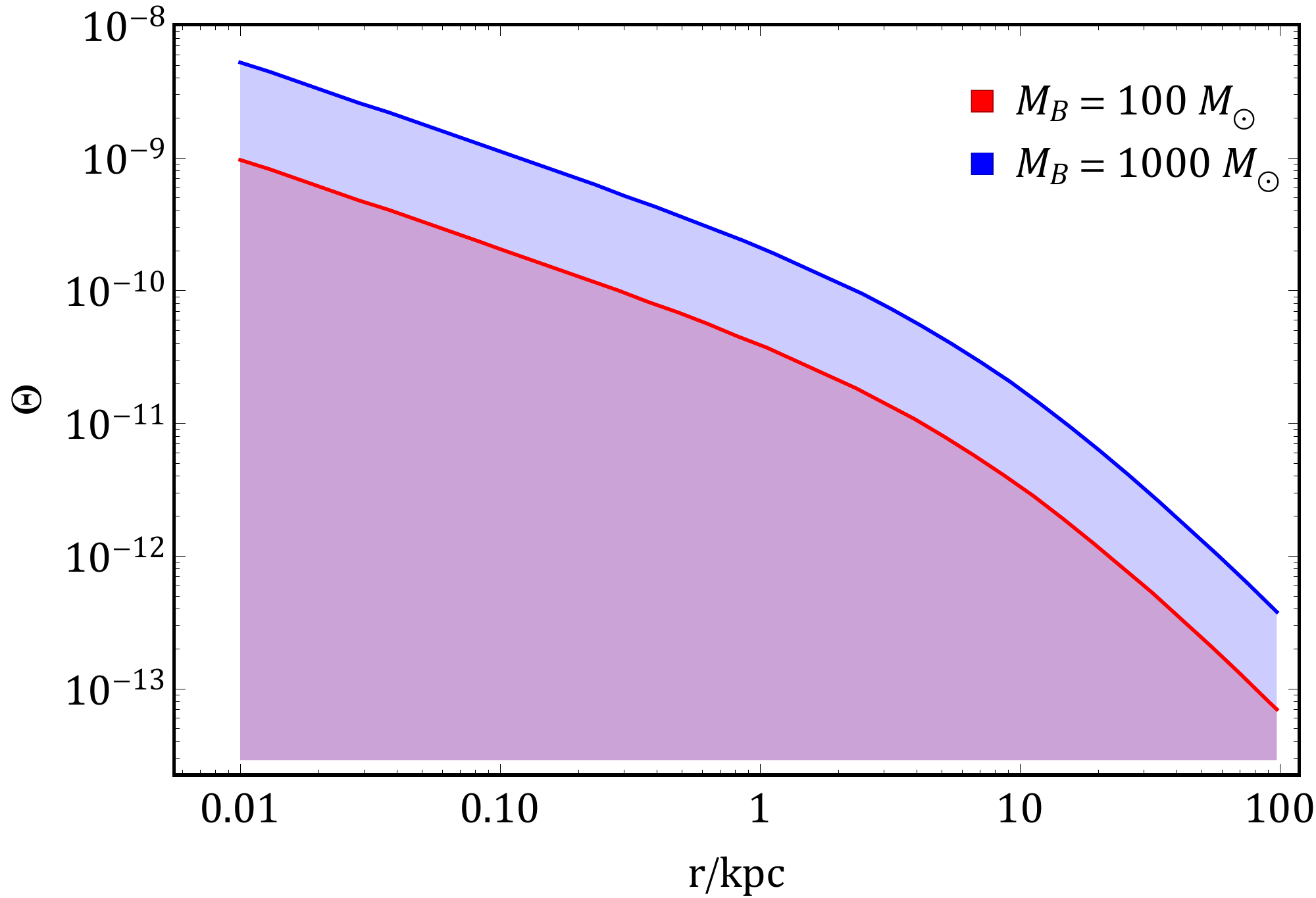}
	\caption{\label{fig:theta_pos}
        The detectable regions for dimensionless temperature $\Theta$ with different locations of PSR-BH binaries. We assume the pulsar mass is $1.6 \, M_\odot$.  The red (blue) shadow regions represent the result from the mass of BH is $100 \, M_\odot$ ($1000 \, M_\odot$).
	}
\end{figure}

For a larger detectable $\Theta$ range, we can extend the location of the PSR-BH binary to a position with the higher DM density, where the mass accretion rate is effectively enhanced, hence increases the orbital phase shift of the binary and enlarges the detectable window of $\Theta$. We consider the PSR-BH binary with $0.01 \, \mathrm{Hz}$ GW frequency, circular orbit and $1 \, \mathrm{ms}$ pulse period. Within a $10 \, \mathrm{years}$ observation, the detectable range for $\Theta$ is shown in Fig.~\ref{fig:theta_pos}.
It clearly shows a high DM density location near the center of the Milky Way can extend the detectable range of $\Theta$, and the accretion effect on a $1000 \, M_\odot$ BH can extend the range of $\Theta$ up to a value close to $10^{-8}$, which is an upper bound of $\Theta$ inside the Milky Way \cite{Avelino:2015dwa}.

\subsection{Ultralight Dark Matter}\label{uldm_acc_in_bin}

To estimate the accretion rate of ultralight DM, we use Eq.~\eqref{eq:ultralightDM}, which gives the mass accretion rate by a non-rotating BH of mass $M_\mathrm{B}$ traveling through a uniform distributed scalar field. Due to the accretion rate of ultralight DM is relatively weak compared with WIMPs, we mainly focus its accretion at the center of the Milky Way. Follow \cite{Hui:2016ltb}, we apply the central density and virial velocity of the soliton in Eq.~\eqref{eq:ultralightDM}, which gives its accretion rate,
\begin{align}\label{eq:soliton_acc}
	\frac{\dd M_{\rm B}}{\dd t} = 
	\frac{ 2.5 \, M_\odot } {10^{17} \, {\rm yr} } \left( \frac{M_{\rm B} }{\tilde{M}_\mathrm{B} }  \right)^2 \left( \frac{m_{\rm ul}}{\tilde{m}_\mathrm{ul}} \right)^6 \left( \frac{M_{\rm sol}} {\tilde{M}_\mathrm{sol}} \right)^4~,
\end{align}
where the reference BH mass $\tilde{M}_\mathrm{B} = 100 \, M_\odot$, the reference ultralight DM mass $\tilde{m}_\mathrm{ul} = 10^{-22} \, \mathrm{eV}$ and the reference soliton mass $\tilde{M}_\mathrm{sol} = 10^{10} \, M_\odot$. The accretion rate of ultralight DM $\dot{M}_\mathrm{B} \propto M_\mathrm{B}^2$, which is similar with the WIMPs' accretion rate, therefore, in a given parameter setting $(M_\mathrm{B}, m_\mathrm{ul}, M_\mathrm{sol})$, the time evolution of orbital phase shift is similar like the behavior in Fig.~\ref{fig:orbital_phase}.

\begin{figure}[htbp] \centering
	\includegraphics[width=8cm]{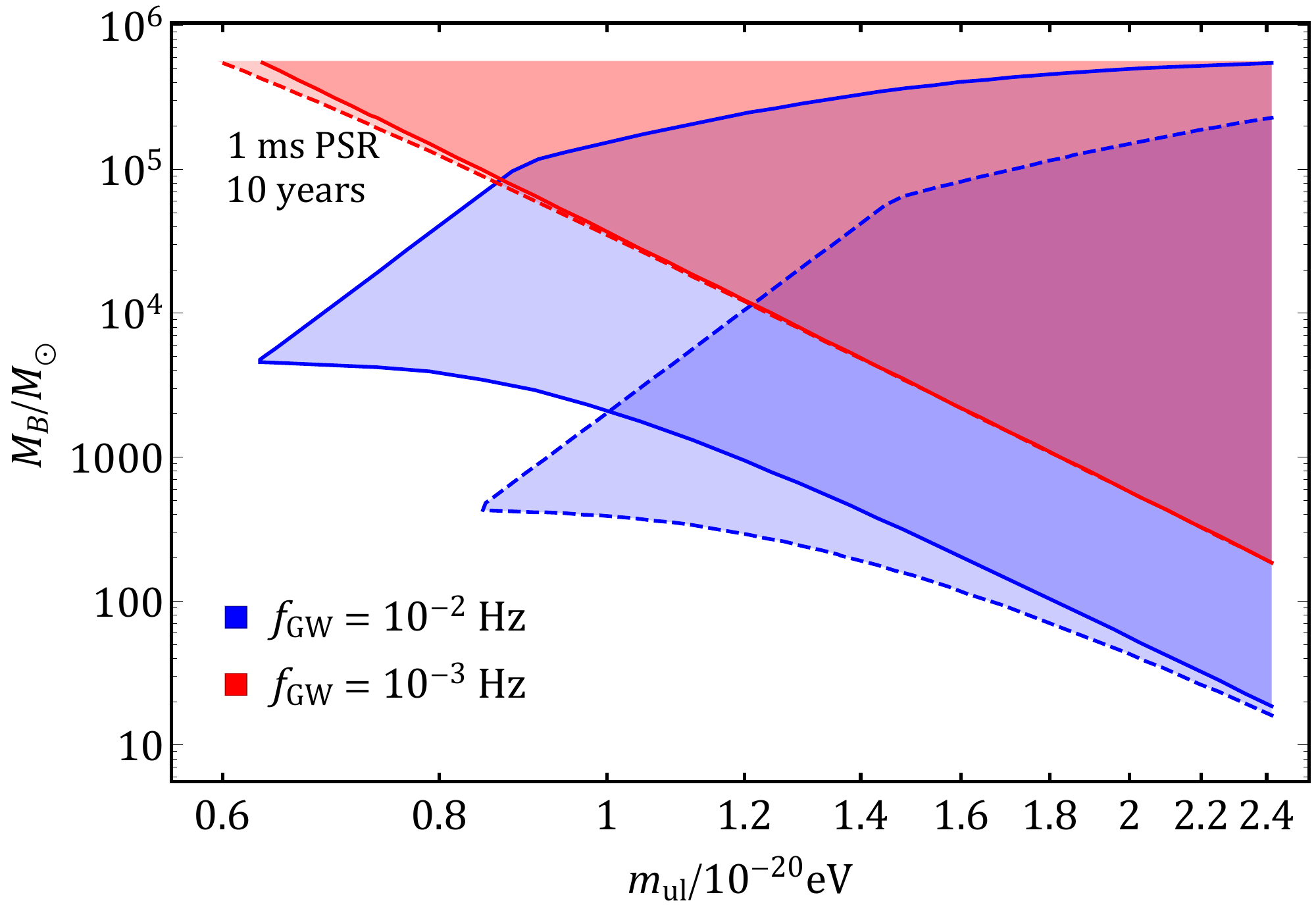}
	\caption{\label{fig:mass_mev}
		The detectable regions for the mass of ultralight DM $m_\mathrm{ul}$ with different BH masses in PSR-BH binaries. The blue (red) shadow regions represent the result from PSR-BH binaries with GW frequency $10^{-2} \, \mathrm{Hz}$ ($10^{-3} \, \mathrm{Hz}$). The solid (dashed) regions represent the result from PSR-BH binaries with eccentricity $e = 0$ ($e = 0.6$).
	}
\end{figure}

In order to find a detectable mass range of ultralight DM in PSR-BH binaries, we set the soliton mass of the Milky Way is $10^9 \, M_\odot$ \cite{DeMartino:2018zkx}, and we follow the procedures introduced in Sec.~\ref{Binaries}, using Eq.~\eqref{eq:require}, where we assume the pulsar mass is $1.6 \, M_\odot$, pulse period is $1 \, \mathrm{ms}$ and the duty time of radio telescope is set as $10 \, \mathrm{years}$. The result is shown in Fig.~\ref{fig:mass_mev}.
We can find that a larger BH mass and higher orbital frequency can extend the detectable mass range of ultralight DM as previous discussions in WIMPs. And the orbital phase shift induced by a larger ultralight DM mass could be detected in PSR-BH binaries with a wider BH mass range, this is because a larger ultralight DM mass can effectively enhance the accretion rate of ultralight DM in Eq.~\eqref{eq:soliton_acc}. Also, contour in Fig.~\ref{fig:mass_mev} is similar with that in Fig.~\ref{fig:mass_theta}, the difference is only a rescale on the values of x-axis. The reason is that the accretion rate in WIMPs and ultralight DM can be generalized as $\dot{M}_\mathrm{B} = f(\alpha) M_\mathrm{B}^2$, where $\alpha$ is the parameter of DM models, namely, $\Theta$ in WIMPs and $m_\mathrm{ul}$ in ultralight DM. For each value of $f(\alpha)$, it relates with two different values in $\Theta$ and $m_\mathrm{ul}$ and corresponds with a specific accretion rate. Meanwhile, this accretion rate determines a BH mass range in PSR-BH binaries, which is the same BH mass range in two DM models, it causes a rescaled x-axis in their contours.

\subsection{Primordial Black Holes}\label{PBH_acc_in_bin}

In order to estimate the magnitude of PBHs accretion rate, we can compare PBHs accretion rate in Eq.~\eqref{eq:PBH_acc} with WIMPs accretion rate in Eq.~\eqref{eq:bondi_formula}. The parameters $\lambda_B$, $\gamma$ are around $\mathcal{O}(1)$ in Eq.~\eqref{eq:bondi_formula}, then we have
\begin{align}
    \frac{\dot{M}_\mathrm{B}^\mathrm{P}}{\dot{M}_\mathrm{B}^\mathrm{W}} \simeq \frac{27}{4}\frac{v}{c} \Theta^{3/2}~.
\end{align}
Here, we use the velocity in rotation curve of the Milky Way \cite{Bhattacharjee:2013exa} to approximate the relative velocity between the BH and PBHs $v$, which is around $200 \, \mathrm{km/s}$. And we set $\Theta \sim 10^{-10}$, which is the largest detectable value in a $10^{-2} \, \mathrm{Hz}$ PSR-BH binary. Then, the accretion rate ratio between PBHs model and WIMPs can be evaluated as $\dot{M}_\mathrm{B}^\mathrm{P}/\dot{M}_\mathrm{B}^\mathrm{W} \sim \mathcal{O}(10^{-17})$. Such a small PBHs accretion rate can hardly be detected in our PSR-BH binary systems. 

Actually, the accretion rate in Eq.~\eqref{eq:PBH_acc} is a time averaged accretion rate, and the mass increment due to accreting PBHs only occur when the astrophysical BH interacts with a PBH. This interaction happens within a very short time interval, so a physical PBH accretion rate can be approximated as a summation of Dirac delta functions 
\begin{align}
	\frac{\dd M_\mathrm{B}}{\dd t} = M_\mathrm{PBH} \sum_{n=1}^{\infty} \delta(t-nt_f)~,
\end{align}
where $t_f$ is the mean free time of each merger of the accreting BH and PBH. Therefore, this mean free time cannot be too long that such a merger event would not happen once during the observational duty time. We can give a brief estimation on this mean free time by dividing the mean free path by the relative velocity between the accreting BH and PBHs. In Eq.~\eqref{eq:mean_free_path}, we assume the mass of accreting BH is $100 \, M_\odot$, PBH mass is $1 \, M_\odot$ and $\rho_\mathrm{DM} = 0.013 \, M_\odot/\mathrm{pc}^{3}$, which is the DM density at the location of the sun in NFW model \cite{Lin:2019yux}. As above, the relative velocity between the accreting BH and PBHs is approximated by the velocity in rotation curve $v \sim 200 \, \mathrm{km/s}$. Then a mean free time is $t_f = l_f/v \sim \mathcal{O}(10^{26}) \, \mathrm{years}$, which is too long to be detected in a PSR-BH binary.

\section{Conclusions}\label{Conclusions}

To summarize, we propose a DM probe that can be used to study the nature of DM. This probe is based on the DM accretion in PSR-BH binaries. The DM accretion could slowly increase the BH mass, and such a mass increment in a PSR-BH binary would affect its gravitational radiation and decrease gravitational potential energy, then induce an orbital phase shift in the orbital evolution of the PSR-BH binary. Various DM models contribute different mass accretion rates and induce distinct orbital phase shifts. After a long observation time in pulse timing, such an orbital phase shift could be detected. From observable orbital phase shifts, it can help us distinguish the DM models and constrain the parameters in DM models. 

In this work, we mainly focus on three DM models, WIMPs, ultralight DM, and PBHs. For WIMPs, the accretion rate follows $\dot{M}_\mathrm{B} \sim M_\mathrm{B}^2 \rho_\mathrm{DM} \Theta^{-3/2}$, a larger BH mass, the higher DM density and a lower dimensionless temperature, all can enhance the WIMPs accretion rate.  Consider WIMPs with $\Theta \sim \mathcal{O}(10^{-12})$, accreted by a PSR-BH binary with a $100 \, M_\odot$ BH mass at the position $10 \, \mathrm{kpc}$ away from the center of the Milky Way, the detectable timescale of induced orbital phase shift is $\mathcal{O}(10) \, \mathrm{years}$. Within a $10 \, \mathrm{years}$ observation,  a value of $\Theta$ can be detected up to $10^{-8}$ in a PSR-BH binary with $10^{-2} \, \mathrm{Hz}$ GW frequency, and the observable range of $\Theta$ can be extended with a larger BH mass and higher orbital frequency. For ultralight DM, we mainly consider the accretion inside the soliton of the Milky Way, due to their weak accretion rate, which follows $\dot{M}_\mathrm{B} \sim M_\mathrm{B}^2 m_\mathrm{ul}^6$, so a larger mass of ultralight DM can effectively improve the detectability of accretion effect. In our parameter setting, the orbital phase shift induced by the accretion of ultralight DM with mass above $\mathcal{O}(10^{-20}) \, \mathrm{eV}$ can be detected, which could help constrain the mass the ultralight DM. For PBHs, their number density inside the Milky Way is small due to a large PBH mass, and this small number density causes an extremely long mean free time of a BH encountering PBHs, around $\mathcal{O}(10^{26}) \, \mathrm{years}$, which is not a detectable timescale. Therefore, a null detection result may indicate a possibility of PBHs as DM or some other DM models with undetected parameter regions.

In above discussions, the mass accretion rate in different DM models basically follows $\dot{M}_\mathrm{B} = f(\alpha) M_\mathrm{B}^2$, where $\alpha$ is the DM parameter. Therefore, a detected DM accretion rate may correspond to different DM parameters in their models, which could cause difficulty in distinguishing them. In pining down a specific DM model, some other constraints on DM \cite{Billard:2021uyg} can be used in breaking this DM model degeneracy, then a DM parameter in this model could be constrained from observed accretion rate. In addition, some other effects such as baryonic matter accretion and dynamical friction in a DM density spike \cite{Chan:2022gqd}, should be taken into consideration in a real data analysis, which is neglected in above calculations.

Apart from mass accretion, some other phenomena could also change the BH mass and induce an orbital phase shift in PSR-BH binaries, such as superradiance effect around a Kerr BH \cite{1971JETPL..14..180Z, 1972JETP...35.1085Z, Brito:2015oca}, which could extract $\mathcal{O}(10\%)$ mass of host BH to form a light boson cloud \cite{Ng:2020ruv, Hui:2022sri}. Therefore, detailed studies on these significant effects in PSR-BH binaries could shed light on unknown physics and show a great potential of the PSR-BH binary.

\section*{Acknowledgements}
We would like to thank Lam Hui, Yi Wang, Henry Tye, and Leonardo Modesto for the very helpful comments and advice.

\begin{appendix} 

\section{Weakly Interacting Massive Particles Accretion}\label{WIMPAppendix}

Here we will introduce the Bondi accretion formula which applies to the WIMPs. We assume spherical symmetry, and a steady state accretion, with the uniform density $\rho_\infty$ and pressure $p_\infty$ at spatial infinity. Thus, at infinity, the sound speed $c_s$ reads,   $c_{s,\infty} = (\frac{\gamma p_\infty}{\rho_\infty})^\frac{1}{2}$. The steady flow translates as
\be  
\dot M = - 4 \pi r^2 \rho u~,
\ee
where $u$ is the radial velocity. 
We can then write the equations for momentum conservation,
\be \label{momentum1}
u \frac{\dd u}{\dd r } + \frac{c_s^2}{\rho} \frac{\dd \rho } {\dd r} + \frac{G M}{r^2 } = 0~,
\ee
and the mass conservation,
\be 
\frac{1}{\rho} \frac{\dd \rho}{\dd r } = -\frac{2}{r} - \frac{1}{u} \frac{\dd u}{\dd r}~. 
\ee
Substituting the latter into the former yields, 
\be  
\label{BondiEq} \frac{1}{2} \left( 1 - \frac{c_s^2}{u^2} \right) \frac{\dd u^2}{\dd r} = \frac{- G M }{r^2} \left(1- \frac{2 c_s^2 r}{G M} \right)~,
\ee
which is the Bondi equation. 
We can contemplate the term between parentheses on the LHS. As $r \to \infty $ and $c_s \to c_{s,\infty}$ that term is negative. Whereas for $r \to 0$ it is positive again. Assuming continuity, there must be a point where it vanishes.  At that point the LHS vanishes simultaneously for the $u^2(r_s) \equiv u_s^2  = c_s ^2 $. 
For our case, the physically relevant solution is this so called transonic solution with the condition
\be 
u^2 \to 0 \, \, \,  {\rm as } \, \, \, r \to \infty~.
\ee
From equation (\ref{BondiEq}), the sonic point is at 
\be 
r_s = \frac{GM}{2 c_s^2}~.
\ee
Moreover, from equation (\ref{momentum1}), with the Polytropic condition 
\be 
P \sim \rho^\gamma~,
\ee 
one simply derives
\be 
\frac{u^2}{2} + \frac{c_s^2(r)}{\gamma - 1}  - \frac{GM}{r} = \frac{c_{s,\infty}^2 }{\gamma - 1}~.
\ee
Then, evaluating this at the sonic point, we find,
\be  
c_s(r_s) = c_{s,\infty} \sqrt{\frac{2}{5 - 3 \gamma}}~.
\ee
This finally brings us to the Bondi accretion law,
\be \label{FinalBondi}
\dot{M}_B = 4 \pi \lambda_B (G M )^2 \frac{\rho_\infty}{ c_{s,\infty}^3}~,
\ee
with 
\be 
\lambda_B = \frac{1}{4} \left(\frac{2}{5 - 3 \gamma} \right)^{\frac{5-3 \gamma}{2 (\gamma - 1)}}~.
\ee

Another way of expressing Eq.(\ref{FinalBondi}), as a function of the temperature per particle mass, is using 
\be 
\frac{k_B T}{ m} = \frac{P}{\rho } =  \frac{c_s^2}{\gamma}~,
\ee
where $k_B$ is the Boltzman constant and $m$ is the WIMP mass.
This, inserted in (\ref{FinalBondi}), yields
\be  
\dot{M}_B = 4 \pi \lambda_B (G M )^2 \frac{\rho_\infty m^{\frac{3}{2} } } { \gamma^{\frac{1}{2}} \, (k_B T) ^{\frac{3}{2} } }~,
\ee
Which is equivalent to Eq.(\ref{eq:bondi_formula}).

\section{Hot Particle Dark Matter Accretion}\label{HotAppendix}

We start from the continuity equation, and the vanishing of the divergence of the stress-energy tensor, 
\be 
\label{mass} \nabla_{ \mu} \left(\rho u^{\mu } \right) = 0~, \\
\label{momentum} \nabla_\mu T^{\mu \nu }  = 0~.
\ee
Here, $U^\mu$ is the DM fluid 4-velocity, and $T^{\mu \nu} = \rho (1 + h ) u^\mu u^\nu + p g^{\mu \nu}$.
We assume a Schwarzschild metric background,
\be 
\dd s^2 = - (1- \frac{2M}{r}) \dd t^2 + \frac{1}{1- \frac{2M}{r}} \dd r^2 + r^2 \dd \Omega ^2~.
\ee
The corresponding metric determinant square root is thus 
\be  
\sqrt{-g} = r^2 \sin \theta~,
\ee
and the four-velocity, defined in our spherically symmetric case as 
\be
u^\mu = \frac{\dd x^\mu}{\dd \tau} = (u^t , u^r, 0 ,0)~,
\ee 
with $u^\mu u_{\mu} = -1 $.
The components of the four-velocity thus relate to each other as 
\be 
&&g_{tt} (u^t)^2 + g_{rr} (u_r)^2 = -1 \\ 
&&(u^t)^2 = \frac{ 1- \frac{ 2M}{r}  + (u^r)^2 }{ (1 - \frac{2M}{r} )^2 }~.
\ee
Now we go back to Eq.(\ref{mass}) which yields,
\be  
\frac{\dd}{\dd r} (r^2 \rho u^r) = 0~,
\ee
and Eq.(\ref{momentum}) yields,
\be  
\frac{\dd}{\dd r} \left(r^2 \rho (1 + \frac{\gamma}{\gamma-1} \Theta ) u^t u^r\right) = 0~.
\ee
Once integrated they reduce to
\be 
\dot M = 4 \pi r^2 \rho u^r~,
\ee
and
\be 
(1 + \frac{\gamma  \Theta}{\gamma-1} ) \left( 1- \frac{2M}{r} + |u^r|^2  \right)  =  (1 + \frac{\gamma \Theta_\infty}{\gamma-1}  )~.
\ee
Combining the two equations, and following the same procedure as in Appendix.~\ref{WIMPAppendix} yields Eq. (\ref{Michel}).

\section{Ultralight Dark Matter Accretion}\label{uldmAppendix}

In \cite{Unruh:1976fm}, Unruh considered the Klein-Gordon equation on a Schwarzschild BH geometry,
\be 
g^{\mu \nu} \phi_{, \mu ; \nu} + m^2 \phi= 0~.
\ee
With the separation of variables 
\be 
\phi(t,r,\theta,\varphi) = {\rm e } ^{- i \omega t} f_{\omega l } (r) Y_{l m } (\theta , \varphi)~,
\ee
one is left with,
\be 
\! \! \! \! \left[ \frac{ \rho^2}{r^2} \frac{\dd }{\dd r} (r^2 \rho^2 \frac{\dd}{\dd r}) \!  +  \! \omega^2 \!  -  \!  \rho^2 \!  \left(  \!  m^2  \! \! +  \! \frac{l ( l+1) }{r^2} \!  \right)  \!   \right] \! \!  f_{\omega l } (r) = 0~,
\ee 
where $\rho = (1 - \frac{2M}{r})^\frac{1}{2} $.
The equation is not solvable analytically as it is, but, it can be approximated in three phases, far from the BH, at intermediate distance, and very close. Each of these cases can be analytically solved, then matched by imposing reasonable boundary conditions at infinity and matching the boundary conditions until reaching the BH surrounding.
Having the solution near the BH, one can compute the current term
\be 
J^{\mu} = \frac{1}{2} i \phi^{*} \overleftrightarrow{ \partial} ^{\mu} \phi ~.
\ee
The number of particles absorbed by the BH will be a surface integral of the radial current 
\be 
N = - \int_\mathcal S \sqrt{- g} J^r \dd \mathcal S~.
\ee
Plugging the solution in we get the cross section 
\be 
\sigma(\omega, v) = \frac{N}{\omega v} = \frac{(4 \pi M)^2 (1+ v^2) 2 M \omega }{v^2(1- {\rm e}^{- \frac{ 2 \pi  M \omega (1+ v^2)} {v}})}~.
\ee

\section{Gravitational Potential Energy Changing Rate}\label{GPAppendix}

The gravitational potential energy in a PSR-BH binary system can be estimated as
\begin{align}
	E_p = -\frac{G m_\mathrm{p} M_\mathrm{B}}{r}~.
\end{align}
Due to the DM accretion into BH, the gravitational potential energy is transferred from the DM to the BH, which results in the increasing in BH mass and decreasing in gravitational potential energy, as follows,
\begin{align}\label{eq:GP_change}
	\frac{\dd E_p}{\dd t} = -\frac{G m_\mathrm{p}}{r} \frac{\dd M_\mathrm{B}}{\dd t}~.
\end{align}
During the orbital motion of the binary, the distance $r$ between two components can be expressed in polar coordinate as follows,
\begin{align}
	r = \frac{a(1 - e^2)}{1 + e \cos \psi}~.
\end{align}
Then Eq.~\eqref{eq:GP_change} can be written as
\begin{align}
	\frac{\dd E_p}{\dd t} = -G m_\mathrm{p} \frac{\dd M_\mathrm{B}}{\dd t} \frac{1 + e \cos \psi}{a(1-e^2)}~.
\end{align}
The average gravitational potential energy changing rate over one period is
\begin{align}\label{eq:ave_EP}\nonumber
	\left \langle \frac{\dd E_p}{\dd t} \right \rangle &= -\frac{1}{T} \int_0^T G m_\mathrm{p} \frac{\dd M_\mathrm{B}}{\dd t} \frac{1 + e \cos \psi}{a(1-e^2)} \dd t\\ 
        &= -\frac{1}{T} \int_0^{2 \pi} G m_\mathrm{p} \frac{\dd M_\mathrm{B}}{\dd t} \frac{1 + e \cos \psi}{a(1-e^2)} \frac{\dd \psi}{\dot{\psi}}~.
\end{align}
Here $T$ is orbital period and $\dot{\psi}$ is the angular velocity, which follow
\begin{align}\nonumber
	T &= \frac{2 \pi a^{3/2}}{G^{1/2}(m_\mathrm{p}+M_\mathrm{B})^{1/2}}~,\\
        \dot{\psi} &= \frac{[G(m_\mathrm{p} + M_\mathrm{B}) a (1 - e^2)]^{1/2}}{r^2}~.
\end{align}
Then, the integral Eq.~\eqref{eq:ave_EP} gives the result
\begin{align}
	\left \langle \frac{\dd E_p}{\dd t} \right \rangle = -\frac{G m_\mathrm{p}}{a}\frac{\dd M_\mathrm{B}}{\dd t}~.
\end{align}

\end{appendix}

\end{document}